\documentclass{article}
\usepackage{amsmath, amssymb}

\begin{document}

{\centering

\LARGE Implementation of Motzkin-Burger algorithm \\ in Maple

\vspace{\baselineskip}

\large P.A. Burovsky \\

\vspace{\baselineskip}

burp@front.ru

}

\vspace{\baselineskip}

\begin{abstract}
Subject of this paper is an implementation of a well-known Motzkin-Burger
algorithm, which solves the problem of finding the full set of solutions of
a system of linear homogeneous inequalities. There exist a number of
implementations of this algorithm, but there was no one in Maple, to
the best of the author's knowledge.
\end{abstract}

\section{The problem}

Consider a linear homogeneous system of inequalities with rational 
coefficients
\begin{equation}\label{linear_system1}
l_j(x) = a_{j1} x_1 + \ldots + a_{jn} x_n \leqslant 0,
    \quad a_{ji} \in \mathbb{Q}, \quad (j=1,\ldots,m).
\end{equation}
Let $r$ be the rank of the matrix of \eqref{linear_system1}.

Recall some common facts from the theory of convex sets (see for example, 
\cite{Fulton}). Every solution set $X_1,\ldots,X_p$ of a system 
\eqref{linear_system1} generates a convex cone $C$ of solutions
\begin{equation}\label{decomposition}
  k_1 X_1 + \ldots + k_p X_p, \qquad k_i \geqslant 0.
\end{equation}

We call a finite set of solutions $X_1,\ldots,X_p \in \mathbb{Q}^n$ a set of 
generators for $C$ if every element of $C$ is a conical combination
\eqref{decomposition} of $X_1,\ldots,X_p$. Hereinafter we will consider the
minimal sets of generators only, i.e. such that none of the $X_i, i=1,\ldots,p$
is expressed from the others with positive coefficients. We call $d$ the 
dimension of a cone $C$ if $d$ is the dimension of the minimal linear space 
spanned by $C$.

A solution locus of each inequality is a half-space of $\mathbb{Q}^n$. 
For any subsystem $\{l_j(x)\leqslant 0, j \in J\}$
of \eqref{linear_system1} its solution set is a convex polyhedral
cone. The faces of the latter are the intersections of $C$ with solution loci 
of some equation subsystems $\{l_j(x)=0, j \in I \subseteq J\}$ with
linearly independent left-hand sides.

The problem discussed in this paper is to solve \eqref{linear_system1} a 
over $\mathbb{Q}^n$. This problem is closely related to the following one.
Extend the system \eqref{linear_system1} by a new inequality
\begin{equation}\label{new}
   l(x) = a_{1} x_1 + \ldots + a_{n} x_n \leqslant 0.
\end{equation}
Not all the points of the cone $C$ satisfy \eqref{new}, thus the system
\eqref{linear_system1} together with \eqref{new} define a new cone $C^\star$.
We are interested in computing the set of generators for $C^\star$ given the
set of generators for $C$. The Motzkin-Burger algorithm solves this problem.

The iteration of the Motzkin-Burger algorithm solves the first problem
by means of extending the trivial system $\{0\leqslant 0\}$ by the 
inequalities of \eqref{linear_system1} one \linebreak by one.

\section{Algorithm description}

\subsection{Theoretical study}

Let us recall some well-known facts (generally following
\cite{Fulton}, \cite{Chernikov}, \cite{Solodovnikov}).

The set of generators for a cone $C^\star$ consists of generators
for cone $C$ (which satisfy the inequality $l(x) < 0$) and
the whole set of generators for cone $C \cap H$ where $H:=\{l(x)=0\}$.
Thus, in order to compute the cone $C^\star$ we need to study both
the structure of the cone $C$ and structure of the intersection.

The cone $C$ is a Minkovsky sum of its linear subspace $L$ and the strongly 
convex cone~$P$ (that is the cone with apex at the origin and contain no line
through the origin). Vectors $u$ of the base $U$ of $L$ 
satisfy $\{l_j(u)=0, j=1,\ldots,m\}$, whereas any element $v$ of
set $V$ of generators of~$P$ satisfies $l_j(v) < 0$ for at least one
$j$.

Let us denote vectors $X$ belonging to $U$ or $V$ by $X^+\,$ if
$\,l(X)>0$, $X^-\,$ if $\,l(X)<0\,$ and $X^0\,$ if $\,l(X)=0\,$.
Also denote the linear subspace of the new cone with its base and the strongly 
convex cone of the new cone with its set of generators
by $L^\star$, $U^\star$, $P^\star$, $V^\star$, respectively. Hence
the vector $u^0$ belongs to $U^\star$, while $u^-$, $v^0$ and
$v^-$ belong to $V^\star$. Although, $\{u \in U : l(u)=0\}$ is not
$U^\star$. By discarding all $u^+$-s and $v^+$-s we lose a number
of generators for the cone $C$ and have to be replaced them
by the whole set of generators for $H \cap C$.

Since $H \cap C = \left(H \cap L \right) \cup \left(H \cap P\right)$
we can describe how to convert $U$ to $U^\star$ and $V$ to $V^\star$
separately.

There are two cases with respect to the intersection $H \cap L$:
$H \cap L = L$ or $H \cap L \not=L$. In the first case, $U^\star =
U$. 

In the second case the new inequality \eqref{new} reduces the
dimension of $L$ by one.
Choose one $u^- \in U$ (if there is no such vector in $U$, take
$-u^+$). Let us form $\dim L - 1$ pairs of vectors $u^-$ and $u_j
\in U\setminus\{u^-\}$. For every pair consider the conic
combinations
\begin{equation}\label{pair's cone}
   au^- + bu_j, \quad a,b \geqslant 0.
\end{equation}
The intersection of $H$ with the cone \eqref{pair's cone} is the
ray of positive multiplies of
\begin{equation}
  u^\star_j := -l(u^-)u_j + l(u_j)u^-.
\end{equation}
It is clear that $l(u_j^\star) = 0$. Thus, $\dim L - 1$ vectors
$u^\star_j$ are the base of $L^\star$. Note that the
representation of the base of $L$ is not unique since we can
choose different vectors for $u^-$ if they exist.

The conversion of $V$ to $V^\star$ also depends on existence of
abovementioned $u^-$. 
The element $u^-$ belongs to $V^\star$ if it exists.
New inequality \eqref{new} induces the affine transformation
\begin{equation}\label{transform}
  -l(u^-)v + l(v)u^-
\end{equation}
of elements $v \in V$ that results in that every vector
\eqref{transform} satisfies \eqref{new}. It is evident that the set
$\{u^-,-l(u^-)v + l(v)u^- \text{ for all } v \in V\}$ is minimal in the
abovementioned sense, thus the latter is the set of generators
$V^\star$.

If there exist no such $u^- \in U$, the transformation \eqref{transform} 
of $V$ is impossible. The new inequality separates $P$ into two parts. As
already was mentioned, the elements $v^-$ belong to $V^\star$.

To find all the vectors that lie in $H$ let us consider any pair
$v^-,v^+ \in V$. For this pair the combination
\begin{equation}
  v^\star_k := -l(v^-)v^+ + l(v^+)v^-
\end{equation}
lies in $H$. The set of all $v^\star_k$-s generates the convex
polyhedral cone in $H$, but there are too many elements of
$\{v^\star_k\}$ for it to be the set of generators for the latter cone.
The minimal faces that contain a pair of generators are exactly 
two-dimensional ones. Therefore, in order to avoid the superfluous 
solutions we only need to reject all the pairs of generators that are not 
lying on faces of dimension $2$.

Checking if two vectors
lie on a face of dimension $2$ may be done in two ways. The first way is to 
check whether there exists $\{l_j(x), j \in J, |J|=r-2\}$ with linear 
independent $l_j(x)$ such that $l_j(v^-)=l_j(v^+)=0$ holds true for all 
$j \in J$. The second way is to check there exist no third $v \in V$ such that 
$l_j(v)=0$ for all $j$ such that $l_j(v^-)=l_j(v^+)=0$ (are not taken into account
the rank $r$ nor linear independence of $l_j$-s).

We therefore come to the algorithm that solves a system of linear inequalities 
over $\mathbb{Q}^n$:

\vspace{\baselineskip}

\begin{description}
\setlength{\itemsep}{-\parsep}
\setlength{\labelsep}{1em}

\item[Input:\ \ ] $S:=\{l_1(x) ,\ldots,l_m(x) \}$ --- the left-hand sides of the 
	inequalities 
\item[\hspace{3.4em}] of \eqref{linear_system1}, n --- the space dimension.
\item[Output:] $U=\{u_1,\ldots,u_t\}$ --- the base of the maximal linear 
	subspace $L$ of 
\item[\hspace{3.4em}] the solution cone of~\eqref{linear_system1},
\item[\hspace{3.4em}] $V=\{v_1,\ldots,v_s\}$ --- the set of generators for the strongly 
	convex 
\item[\hspace{3.4em}] cone $P$ of the solutions of the system~\eqref{linear_system1}.

\end{description}

\begin{enumerate}
\setlength{\itemsep}{-\parsep}
\setlength{\labelsep}{1em}
  \item $U_{current}:=\{(1,\ldots,0),\ldots,(0,\ldots,1)\}$;
  \item $V_{current}:=\emptyset$;
  \item $S_{current}:=\{0 \leqslant 0\}$;
  \item $i:=1$;
  \item $l:=l_i(x)$;
  \item \textbf{if} $\exists u \in U_{current}$: $l(u) \not= 0$, \textbf{then}
    \par\vspace{-0.75\baselineskip}
    \begin{description}
    \setlength{\itemsep}{-\parsep}
    \setlength{\labelsep}{1em}
      \item[\quad] $U_{current} := \{l(u) u_i - l(u_i) u, \quad u_i
            \in U_{current}\}$;
      \item[\quad] $n:=-(l(u)/|l(u)|)u$;
      \item[\quad] $V_{current} := \{n, -l(n) v_i + l(v_i) n , \quad v_i \in 
		V_{current}\}$;
    \end{description}
    \par\vspace{-0.75\baselineskip}
  \item \textbf{else}
    \par\vspace{-0.75\baselineskip}
    \begin{description}
    \setlength{\itemsep}{-\parsep}
    \setlength{\labelsep}{1em}
      \item[\quad] $V^1_{current}:=\{v \in V_{current} | \;
        l(v) \leqslant 0 \}$; $V^2_{current}:=\emptyset$;
      \item[\quad] \textbf{for} $\forall (v_k, v_s) \in
        V_{current}^2 \;:\; l(v_k) < 0$, $l(v_s) > 0$ \textbf{do}
        \begin{description}
        \setlength{\itemsep}{-\parsep}
        \setlength{\labelsep}{1em}
          \item[\quad] $S^\star:=\{ l_j(x) \in S_{current} | \;
            l_j(v_k) = l_j(v_s) = 0\}$;
          \item[\quad] \textbf{if} $S^\star \not= \emptyset$
            \textbf{then}
            \begin{description}
            \setlength{\itemsep}{-\parsep}
            \setlength{\labelsep}{1em}
              \item[\quad] \textbf{for} $\forall\;v \in V_{current}
                \setminus \{v_k,v_s\}$

                \begin{description}
                \setlength{\itemsep}{-\parsep}
                \setlength{\labelsep}{1em}
                  \item[\quad] \textbf{for} $\forall\; l_j(x)
                    \in S^\star$ \textbf{do}

                    \begin{description}
                    \setlength{\itemsep}{-\parsep}
                    \setlength{\labelsep}{1em}
                      \item[\quad] \textbf{if} $l_j(v)
                        \not= 0$ \textbf{then}
                      \item[\qquad]$V^2_{current}:=
                        V^2_{current}\cup\{-l(v_s)v_k+
                        l(v_k)v_s \} $;
                      \item[\quad] \textbf{end if;}
                    \end{description}

                  \item[\quad] \textbf{end do;}
                \end{description}

              \item[\quad] \textbf{end do;}
            \end{description}
          \item[\quad] \textbf{end if;}
        \end{description}
      \item[\quad] \textbf{end do;}
    \end{description}
    \par\vspace{-0.7\baselineskip}\par
    \textbf{end if;}
  \item $V_{current}:=V^1_{current} \cup V^2_{current}$;
  \item $S_{current}:=S_{current} \cup \{l_i(x) \}$;
  \item $S:=S \setminus \{l_i(x)\}$; 
  \item $i:=i+1$;
  \item \textbf{if} $S = \emptyset$ \textbf{then}
    \par\vspace{-0.75\baselineskip}
    \begin{description}
    \setlength{\itemsep}{-\parsep}
    \setlength{\labelsep}{1em}
      \item[\quad] goto $14$;
    \end{description}
    \par\vspace{-0.75\baselineskip}
    \textbf{end if;}
  \item goto $5$;
  \item return $U_{current}, V_{current}$.
\end{enumerate}

\subsection{Some improvements}

There are some tricks concerning the preparation of the system in order to
lower the number of inequalities and/or variables.

Firstly, the system may have no occurrences of some of $x_k$,
$k=1,\ldots,n$. In this case we can ``clean up'' the system and thus
reduce the number of variables.

Secondly, we can avoid the linear subspace $L$ of cone of solutions by
performing the change of variables. This trick also enables one to simplify the
original system of inequalities. 

The dimension of $L$ is $n-r$, where $r$ is the rank of matrix of 
\eqref{linear_system1}, as was mentioned above. Let us define new $r$ variables 
$y_1,\ldots,y_r$ to equal any $r$ linear independent left-hand sides of 
\eqref{linear_system1} (for example, first $r$ ones):
\begin{flalign}
  l_j(x) = - y_j &\leqslant 0, \quad j=1,\ldots,r,\\
  l_j(x) &\leqslant 0, \quad j=r+1,\ldots,m.
\end{flalign}
By solving the system $\{l_j(x) = - y_j, j=1,\ldots,r\}$ for
$\{x_1,\ldots,x_n\}$ we obtain the $(n-r)$-dimensional space of solutions

\begin{align}\label{subs1}
\begin{split}
  &x_{i_1} = f_1(y_1,\ldots,y_r,x_{i_{r+1}},\ldots,x_{i_{n}}),\\
  &\dots\\
  &x_{i_r} = f_r(y_1,\ldots,y_r,x_{i_{r+1}},\ldots,x_{i_{n}}),\\
\end{split}\\
\begin{split}
  &x_{i_{r+1}} = x_{i_{r+1}}\\
  &\dots\\
  &x_{i_n} = x_{i_n}
\end{split}
\end{align}
where $(i_1,\ldots,i_n)$ is the permutation of indices $1,\ldots,n$ (we don't
know \textit{a priori} which variables $x_i$ are most likely to be solved for).

Upon substitution \eqref{subs1} the system \eqref{linear_system1} does not 
depend on $x_{i_{r+1}}, \ldots, x_{i_n}$\, because\, $l_{r+1}(x),\ldots,l_m(x)$
are the linear combinations of the first $r$ ones. Therefore, substitution 
\eqref{subs1} reduces the system to that of $m$ inequalities for $r$ 
independent variables $y_1,\ldots,y_r$.
The $r$ inequalities of reduced system are simplified
to be just $-y_i \leqslant 0$. This inequality subsystem has the
\textit{a priori} known solution $E_r$ --- the $r$-dimensional
linear space base. Hence we may efface these inequalities from the
reduced system. More precisely, solving the system we iterate
Motzkin-Burger algorithm starting from the system of inequalities
$\{-y_i \leqslant 0, i=1,\ldots,r\}$ and $U=\emptyset$ and
$V=E_r$. Thus the system is reduced to of $m-r$ inequalities for
$r$ independent variables. Because $U=\emptyset$ at the start of
algorithm, there is no need in any parts of algorithm except for
the most complicated part that is the conversion of $V$ to
$V^\star$ when there is no $u^-$ exists.

The application of the Motzkin-Burger algorithm yields a number of solutions
of the reduced system. By means of $n-r$ substitutions \eqref{subs1} and
$(x_{i_{r+1}},\ldots,x_{i_{n}}) = E^k_{n-r}$, where $E^k_{n-r}$ are the base
vectors of $(n-r)$-dimensional linear space, we then obtain the solution set
of the original system.

The case of $r = n$ is also of interest for simplifying the system despite that
there is no lowering the number of variables. The system is reduced to that of
$m-n$ inequalities for $n$ independent variables in this case. The calculation
procedure is repeated in full except for that there is no need to consider the
substitution the $(n-r)$-dimensional linear space base vectors to the several
components $(x_{i_{r+1}},\ldots,x_{i_{n}})$ of any solution.

Also of interest is the case of $r=m$ where $m$ is the number of
inequalities. Upon the abovementioned substitution, the system is
reduced to the diagonal one. Therefore, the whole system have
\textit{a priori} known solution in terms of new variables.

We therefore come to the algorithm of solving the system
\eqref{linear_system1} with special preparation prior to the
application of the Motzkin-Burger algorithm:

\vspace{\baselineskip}

\begin{description}
\setlength{\itemsep}{-\parsep}
\setlength{\labelsep}{1em}

\item[Input:\ \ ] $S:=\{l_1(x) ,\ldots,l_m(x) \}$ --- the left-hand sides 
	of the inequalities  
\item[\hspace{3.4em}] of \eqref{linear_system1}, n --- the space dimension.
\item[Output:] $U=\{u_1,\ldots,u_t\}$ --- the base of the maximal linear 
	subspace $L$ of 
\item[\hspace{3.4em}] the solution cone of~\eqref{linear_system1},
\item[\hspace{3.4em}] $V=\{v_1,\ldots,v_s\}$ --- the set of generators for 
	the strongly convex 
\item[\hspace{3.4em}]cone $P$ of the solutions of the system~\eqref{linear_system1}.

\end{description}

\begin{enumerate}
\setlength{\itemsep}{-\parsep}
\setlength{\labelsep}{1em}

\item Find all indices $Bad \subset \{1,\ldots,n\}$ such that $S$ has
        no occurrences of $x_j$,\; $j \in Bad$;
\item Renumber the indices (not encountered in $Bad$) of variables
    so that $S$ explicitly depend on $n-|Bad|$ variables only;
\item Choose $Base$ --- the maximal linear independent subsystem of
    $\{l_j(x), j=1,\ldots,m\}$; $r:=|Base|$;
\item Solve $\{Base_i = -y_i, i=1,\ldots,r\}$ for $\{x_1,\ldots,x_n\}$ to
    obtain set of identities \eqref{subs1};
\item $I := \{i_{r+1},\ldots,i_n\}$;
\item $U^\bullet:=\emptyset$;
\item $V^\bullet:= \{\underbrace{(1,0,\ldots,0)}_{\text{$r$ components}},
    \ldots,(0,\ldots,0,1)\}$;
\item \textbf{if} $r < m$ \textbf{then}
    \par\vspace{-0.75\baselineskip}
\begin{enumerate}
    \setlength{\itemsep}{-\parsep}
    \setlength{\labelsep}{1em}
       \item Substitute \eqref{subs1} to $S$ in order to obtain $S^\bullet$;
       \item Reorder the inequalities $S$ such that $-y_i \leqslant 0$
            are the first $r$ ones.
       \item Iterate the part of Motzkin-Burger algorithm
        mentioned above excluding the items 1, 2, 3, 4, 6, considering
        $U_{current}=U^\bullet$, $V_{current}=V^\bullet$,
        $S_{current}=S^\bullet$, $i = r+1$ as the initial condition.
    \end{enumerate}
    \textbf{end if;}

\item $E:=\{\underbrace{(1,0,\ldots,0)}_{\text{$n-r$ components}},\ldots,
    (0,\ldots,0,1)\}$;
\item For each element $e \in E$ substitute $\{y_1=\ldots=y_r=0\}$ and
    $(x_{I_1},\ldots,\linebreak x_{I_{n-r}})=e$ into \eqref{subs1} and
    form $u := (x_1,\ldots,x_n) \in U$;
\item For each element $v \in V^\bullet$ substitute
    $\{(y_1,\ldots,y_r)=v\}$ and $(x_{I_1}=\ldots=x_{I_{n-r}})=0$ into 
    \eqref{subs1} and form $v := (x_1,\ldots,x_n) \in V$;
\item return $U, V$.
\end{enumerate}

\subsection{Algorithm complexity}

Let us not distinguish arithmetic and comparison operations. Let $n$ be
the number of independent variables, $p$ be the number of elements in $V$
at the current step, $l$ be the number of $v^-, v \in V$, $k$ be the number of
$v^+, v \in V$, $q$ be the number of inequalities already examined by the
current step and $m$ be the number of inequalities in total.

The calculation of the value of $l(x)$ has the complexity $2n-1$.
If there exists such $u \in U$ that $l(u)\not=0$, than the
conversion $U \rightarrow U^\star$ and $V \rightarrow V^\star$
requires the calculation of $n-1$ differences of the kind of
$l(u^-)u_j-l(u_j)u^-$ with the total complexity $2 \cdot 2n + 1 =
4n+1$. Therefore, the overall complexity of converting both lists
is $2(n-1)(4n+1)\sim 8n^2$.

The case when no such $u \in U$ exists is of baffling complexity.
Let us estimate the complexity of one iteration of this case as a function
$f(p,q,n)$. For efficiency reasons we suppose that all the values of $l_j(x) \in S$
for all $v \in V$ are computed beforehand in order to exclude repeated
calculations.

Selection of $v^-$ takes $l$ operations. Next step --- to select the pairs
$v^-,v^+$ --- requires $2kl$ more operations. The number of pairs reaches
the maximum if $k=l=\frac{p}{2}$ so let $kl=\frac{p^2}{4}$ below.

For every pair $v^-,v^+$ we need to choose those inequalities
$l_j(x)$ for which $l_j(v^-)=l_j(v^+)$ is true. This requires $2q$
operations. Then we should examine every pair on whether $p-2$
elements of $V$ do not zero all the chosen inequalities. Let the
number of the chosen inequalities be as large as possible
\textit{i.e.,} $\frac{q}{2}$.
Therefore, checking all the pairs have the complexity
$\frac{p^2}{4}(p-2)\frac{q}{2} = \frac{p^3q}{8} - \frac{p^2q}{4}$.

Let all the pairs satisfy the conditions above, then the overall
complexity for all new the elements $v$ is $\frac{3p^2}{4}$.

In total, $f(p,q,n) = pq(2n-1) + \frac{p}{2} + \frac{p^2}{2} + \frac{p^3q}{8}
- \frac{p^2q}{4} + \frac{3p^2}{4} = \frac{1}{8}p(p^2q+10p-2pq+16qn-8q+4)
\sim \frac{1}{8}p^3q$.

These intensive calculations are possible since $p_0=3$ because up to
$p \leqslant = 2$ there is only one (or no one) pair of generators for $V$. 
Therefore, up to one new generator is produced by algorithm for it to replace
another.

As was mentioned, number $p_k$ of elements $v \in V$ grows as
$\frac{p_{k-1}}{2}+\frac{p_{k-1}^2}{4}$ on every $k$-th step.
Iterating this function one can see that $p_k = p_k(p_{k-1}) =
O(p_0^{2^{k-1}})$, so the worst-case complexity of pure iteration
of the most complicated part of algorithm is $f(p_0,m,n) =
O(mp_0^{(2^{m-4})^3}/8) = O(mp_0^{2^{3(m-4)}}) = O(mp_0^{2^{3m}})
= O(m3^{2^{3m}})$.

Practice shows that, in fact, such a terrible complexity is almost impossible
to happen. There are many pairs rejected on every step. The number of pairs
generally is not too large.
Number of probe inequatilities for every pair usually lower than $\frac{q}{2}$.
Usually systems of full rank are incompatible for sufficiently large number of
inequalities. Overall computation time nevertheless depends on $n$ for the
moderate systems. Despite of it, the data grows exponentially.

\subsection{Practical experience}

We implemented abovementioned algorithms in Maple as one  
package \linebreak \texttt{motzkin\_burger} consists of three user procedures:
\texttt{Conehull}, \texttt{MB} and \linebreak \texttt{CheckSolutions}. The first
of these solves the problem to compute a set of generators for solution cone
of system of inequalities. The second is the single Motzkin-Burger iteration.
Last of these is the tool that allow user to check up the set of solutions on
a correctness. Prototypes of these are

\vspace{\baselineskip}

\texttt{Conehull(L,x,n,options)}

\vspace{\baselineskip}

\texttt{MB(L,U,V,x,n)}

\vspace{\baselineskip}

\texttt{CheckSolutions(L,S,x,n)}

\vspace{\baselineskip}

Here \texttt{L} is a list of homogeneous polynomials of degree $1$
in terms of variables \texttt{x${}_{\text{\texttt{i}}}$}, \texttt{n}
is dimension of space of solutions; \texttt{U} is the base of linear space
of cone of solutions, \texttt{V} is the set of generators of strongly convex 
cone in latter cone; \texttt{S} is any solution set. Parameter \texttt{options},
at this moment, may accept only one value, ``\texttt{as is}'', what gives an 
instruction do not perform change of variables.

Both \texttt{Conehull} and \texttt{MB} return an \texttt{exprseq} that 
consists of two $2$-dimensional lists which are abovementioned lists \texttt{U} 
and \texttt{V}. \texttt{CheckSolutions} returns an \linebreak \texttt{exprseq} too, but
it might by interpret in another way: first is the list of vectors that are
true solutions whereas the second is the list of uncorrect solutions.

The timings presented in the table below are obtained in Maple 7 on computer
based on Duron 700Mhz, 256 RAM. For every combination of ($n$,$m$,$r$)
there was computed random system having these parameters, with sufficiently 
small integer coefficients. Here $n$ is space dimension, $m$ is the number of 
inequalities in the system and $r$ is the rank of the system. Values $t_1$ and 
$t_2$ are times in seconds for solving the latter systems using or not 
optional parameter of \texttt{Conehull} procedure. All these systems are saved 
in attached to paper package codes text files.

\vspace{\baselineskip}

{\centering 

\begin{tabular}[t]{|p{60pt}|p{60pt}|p{30pt}|p{80pt}|p{45pt}|}
\hline
space dimension, $n$ & number of inequalities, $m$ & rank, $r$ & 
	\texttt{Conehull}, option \texttt{"as is"}, $t_1$, sec & \texttt{Conehull}, 
	$t_2$, sec\\
\hline
 5 & 5 & 5 & 0.010 & 0.120\\
\hline
 5 & 7 & 3 & 0.090 & 0.050\\
\hline
 10 & 10 & 10 & 0.130 & 0.261\\
\hline
 10 & 15 & 5 & 0.150 & 0.180\\
\hline
 20 & 20 & 20 & 1.783 & 2.103\\
\hline
 20 & 30 & 10 & 1.512 & 1.021\\
\hline
 30 & 30 & 15 & 2.864 & 2.003\\
\hline
 40 & 40 & 20 & 8.663 & 4.316\\
\hline
 40 & 40 & 30 & 45.326 & 17.775\\
\hline
 50 & 50 & 40 & 89.118 & 1816.392\\
\hline 
 50 & 50 & 45 & 73.766 & 1549.628\\
\hline
\end{tabular}

}
\vspace{\baselineskip}

One can see that \texttt{Conehull} with option ``\texttt{as is}'' 
computes this examples with always increasing time (with rare exceptions). 
Without the option it behave more complicated: for every fixed $n$ and $m$ the 
case of the 
systems of full rank or ``almost'' full rank may be computed more faster than 
in the case of small rank. Of course, the option slows down the performance
(this is the consequence of inefficient Maple \texttt{subs} implementation
we are using) starting from some dimension, but we hope that the systems of 
larger dimension will be computed more faster with this option. Nevertheless,
to improve this performance bottleneck is the main aim of further work.\looseness=-1

\end{document}